\documentclass[10pt, a4paper,twocolumn]{article}

\usepackage{graphicx}
\usepackage{amsmath}
\usepackage{amsfonts}
\usepackage{amssymb}
\usepackage{hyperref}

\begin{document}

\twocolumn[
  \begin{@twocolumnfalse}
\def\lsim{\mathrel{\raise.3ex\hbox{$<$\kern-.75em\lower1ex\hbox{$\sim$}}}}
\def\gsim{\mathrel{\raise.3ex\hbox{$>$\kern-.75em\lower1ex\hbox{$\sim$}}}}
\sf
\centerline{\Huge LHC and the origin of neutrino mass$^{\S}$  }
\vspace{5mm}
\centerline{\large Goran Senjanovi\' c}
\centerline{{\it International Centre for Theoretical Physics, 34100 Trieste, Italy }}
\vspace{5mm}
\centerline{\large\sc Abstract}
\begin{quote}
\small
 It is often said that neutrino mass is a window to a new physics beyond the standard model (SM).  This is 
    certainly true
if neutrinos are Majorana particles since  the SM with Majorana neutrino mass is not a complete theory.  The classical
text-book test of neutrino Majorana mass, the neutrino-less double beta decay depends on the completion, and 
thus cannot probe neutrino mass. As pointed out already more than twenty five years ago, the colliders 
such as Tevatron or LHC offer a hope of probing directly the origin of neutrino Majorana mass through lepton number 
violating production of like sign lepton pairs. I discuss this in the context of all three types of seesaw
mechanism. I then discuss in detail the situation in $L-R$ symmetric theories, which led originally to the seesaw and which incorporate naturally both type I and type II.  A $W_R$ gauge boson with a mass in a few TeV region could 
easily dominate neutrino-less double beta decay, and its discovery at LHC would have spectacular signatures
of parity restoration and lepton number violation.
  At the end I give an example of a predictive $SU(5)$ grand unified theory that results in a hybrid type I and III seesaw with a light fermion triplet below TeV scale.
\end{quote}
  \end{@twocolumnfalse}]
 {
 \renewcommand{\thefootnote}%
   {\fnsymbol{footnote}}
 \footnotetext[4]{Based on plenary talks at Neutrino 08, Christchurch, New Zealand and Physics at LHC - 2008, Split, Croatia.}
}

 \rm
\newpage

\section{Introduction}

We know that neutrinos are massive but light \cite{Strumia:2006db}. If we wish to account for tiny neutrino masses with only the Standard Model (SM) degrees of freedom, 
we need 
Weinberg's \cite{Weinberg:1979sa} $d=5$ effective operator
\begin{eqnarray}
{\mathcal L}=Y_{ij}\frac{L_iHHL_j}{M},\nonumber
\end{eqnarray}
where $L_i$ stands for  left-handed leptonic doublets and  H  for the usual  Higgs doublet (with a vev $v$). This
in turn produces neutrino Majorana mass matrix
\begin{eqnarray}
M_{\nu}= Y\frac{v^2}{M}.\nonumber
\end{eqnarray}

 The non-renormalizable nature of the above operator signals the appearence of new physics through 
  the mass scale $M$. The main consequence is the $\Delta L=2$ violation of lepton number  through
 
 \begin{itemize}
  
\item 
  neutrino-less double beta decay $\beta\beta 0 \nu$ \cite{Racah:1937qq} 
\item 
same sign charged lepton pairs in colliders.

\end{itemize}

While the neutrino-less double beta decay is a text-book probe of Majorana neutrino mass, the like sign lepton
 pair production, although suggested already long time ago \cite{Keung:1983uu},  has only recently received
 wide attention. In what follows I argue that this process may be our best bet in probing directly the origin of neutrino 
 mass.  Due to the lack of space I can cover only the essential points and I cannot do justice to the fast growing 
 literature in the field.  I have tried to be complete in citations, but I am sure I failed, although not on purpose. 
 I apologize in advance for the omission of papers that merit quotation.  This is a short review, not at all a comprehensive
 study of these interesting issues.
 
  If $M$ is huge, there is no hope of direct observation of new physics. It is often said that large $M$ is more natural, 
for then Yukawas do not have to be small.  For example, $M = 10^{13} \text{GeV}- 10^{14} \text{GeV}$ corresponds to $Y$ of order one.
  However,  small Yukawas are natural in a sense of being protected by chiral symmetries
     and anyway most of the SM  Yukawas are small. Furthermore, large ratios of mass scales need fine-tuning, so there
     is nothing more natural about large $M$.
  I adopt the strategy here of keeping  $M$ free and looking  for theoretical predictions, in particular through grand unification.

  In order to get a window to new physics, we need a renormalizable theory of the above effective operator.  In the minimal
  scenario of adding just one new type of particles, 
   there are only three different ways of producing it through the exchange of heavy 

I) fermion singlet (1C , 1W , Y = 0), called right-handed neutrino; type I seesaw \cite{seesaw},

II) bosonic weak triplet (1C , 3W , Y = 2);  the type II seesaw \cite{Magg:1980ut},

III) fermion weak triplet (1C , 3W , Y = 0);  called type III seesaw \cite{Foot:1988aq}
   
\noindent    where C stands for color and W for $SU(2)$ weak quantum numbers.

It is easy to see that all three types of seesaw lead to one and the same $d=5$ operator above.

By itself, seesaw is not more useful than just Weinberg's  operator unless we can reach the scale $M$ or 
have a theory of these singlets and/or triplets.  This is reminiscent of the Fermi effective theory of low energy 
weak interactions: 
saying that the four fermion interactions can be described by the exchange of a $W$ boson is appropriate either
at the scale $M_W$ or if you have a theory of a $W$ boson. This is precisely what the SM  gauge theory
 had achieved by correlating a plethora of low energy ($E \ll M_W$) weak interaction processes. 
 
    It is often said that neutrino  mass is a window to new physics . This is definitely true if it is Majorana for the SM with Majorana 
    neutrino mass is not complete, as manifest from the $d=5$ operator. In the Dirac case, the theory is complete so that the new
    physics is not mandatory. Flavor violation in charged lepton decays  is GIM suppressed by 
    $\Delta m_{\nu}^2/M_W^2$ and is 
     is too tiny to be observed.  Of course, the new physics may emerge
    from a model of these masses, but it is not mandatory by itself.
    
\begin{figure}
\begin{center}
\includegraphics[scale=0.8]{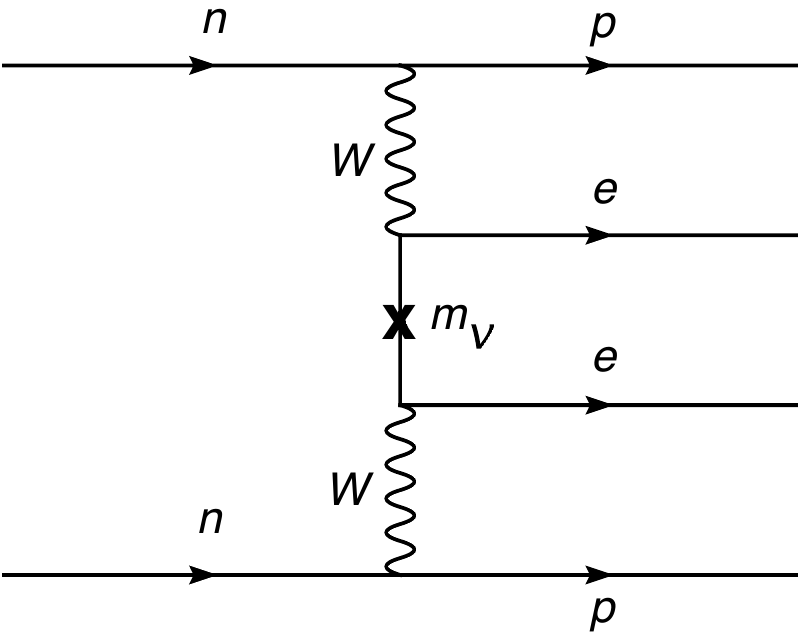} 
\caption{Neutrino-less double beta decay through the neutrino Majorana mass.}\label{bb0v}
\end{center}
\end{figure}

   Majorana case, on the other hand, necessarily  connects $m_\nu$ to new 
    physics, such as desperately searched for  $\beta\beta 0 \nu$, in Fig.\ref{bb0v}.
   This probes neutrino Majorana mass in the range 0.1 - 1 eV.

However, in general $m_\nu$ is not directly connected to $\beta\beta 0 \nu$ decay. While it does produce it,  the inverse is 
not true. $\beta\beta 0 \nu$ decay does not imply the measure of neutrino mass, since it depends on the completion of
the SM needed for the $d=5$ neutrino mass. An example is provided by the $L-R$ symmetric  theory discussed in the next
section.  This is the theory that led originally to the seesaw mechanism, and as such deserves attention. 
As we will see, if the scale of parity restoration is in the few TeV region, the theory offers a rich LHC phenomenology and a
plethora of lepton flavor violating (LFV) processes.

\section{Left-right symmetry and the origin of neutrino mass}
$L-R$ symmetric theories  \cite{leftright} are based on the $SU(2)_L\times SU(2)_R \times U(1)$ gauge group augmented
by parity or charge conjugation.  Then:

\begin{itemize}
\item

$W_L$ implies $W_R$,

\item
$\nu_L$   implies $\nu_R$, with $m_{\nu_R}$ of order $M_R$ through the breaking of $L-R$ symmetry,

\item
Type I seesaw:  connects neutrino  mass to the scale of parity restoration.

\end{itemize}

\begin{figure}[h!]
\begin{center}
\includegraphics[scale=0.8]{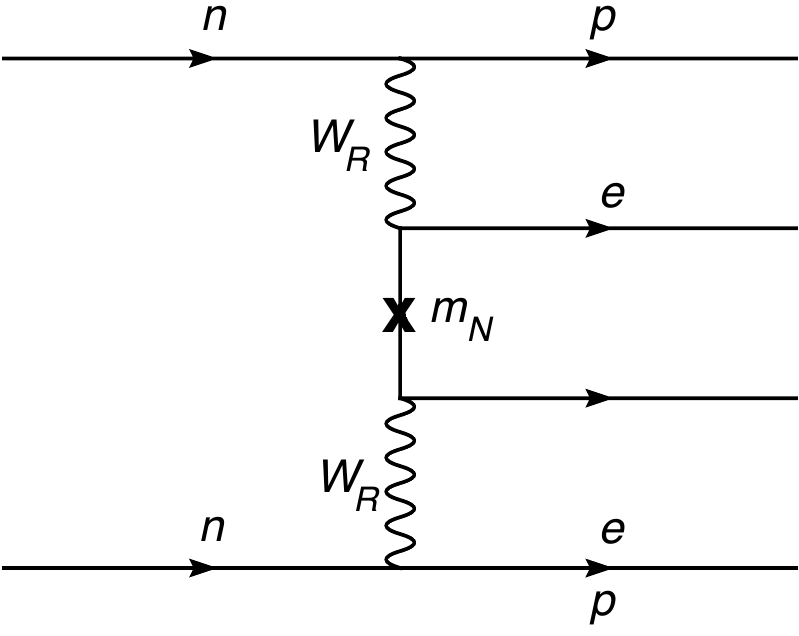} 
\caption{Neutrino-less double beta decay induced by the right-handed gauge boson and right-handed neutrino.}
\label{bb0v2}
\end{center}
\end{figure}

\begin{figure}[h!]
\begin{center}
\includegraphics[scale=0.55]{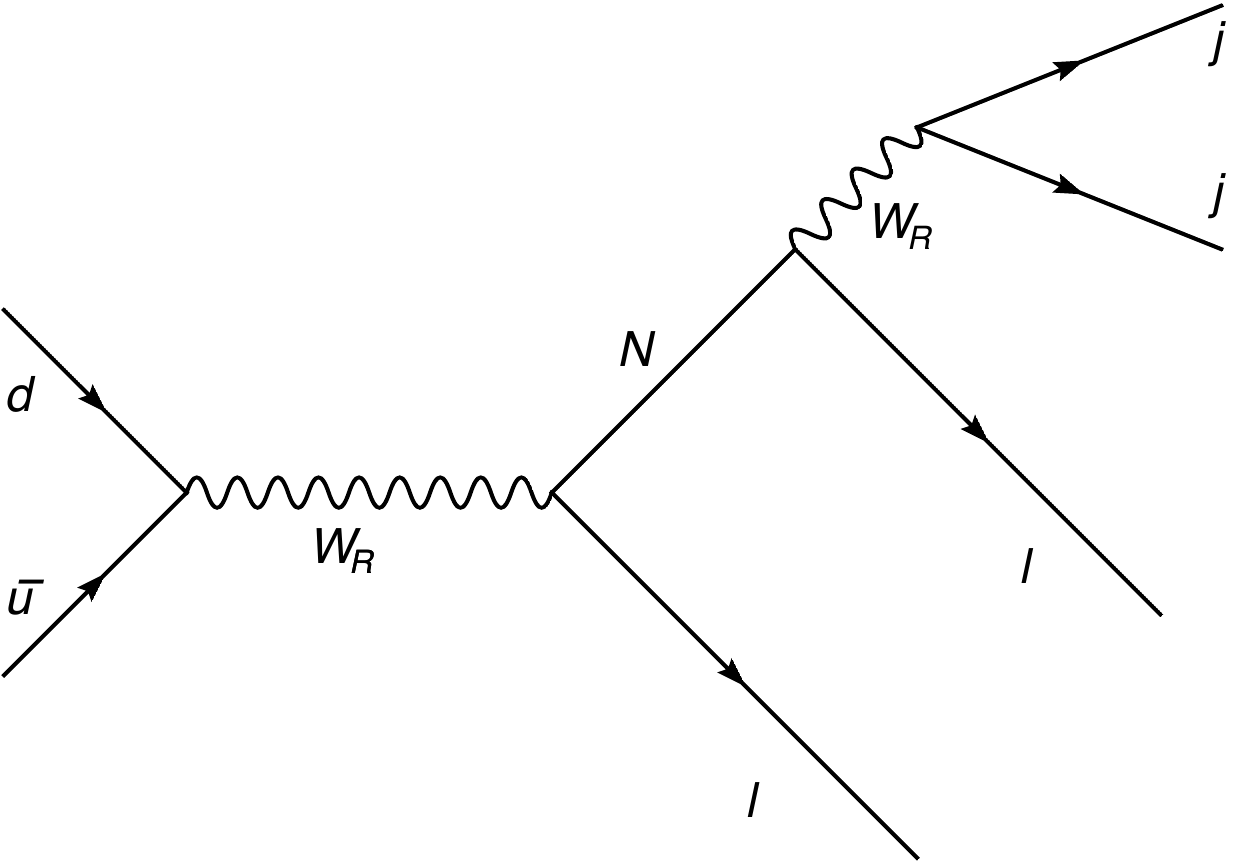} 
\caption{The production of $W_R$ and the subsequent decay into same sign leptons and two jets through
the Majorana character of the right-handed neutrino.}
\label{wr}
\end{center}
\end{figure}

These facts lead immediately to the new contribution to the neutrino-less double beta decay mentioned above,
see Fig. \ref{bb0v2}.
With $W_R$ in the TeV
region and the right-handed neutrino mass $m_N$ in the 100 GeV -TeV region, 
this contribution can easily dominate over
the left-handed one. Neutrino mass can even go to zero (vanishing Dirac Yukawa) while keeping the $W_R$ contribution 
finite. 
\begin{itemize}
\item
Colliders:  produce $W_R$ through Drell-Yan as in Fig. \ref{wr}.
\end{itemize}

\begin{figure*}
\begin{center}
\includegraphics[scale=0.6]{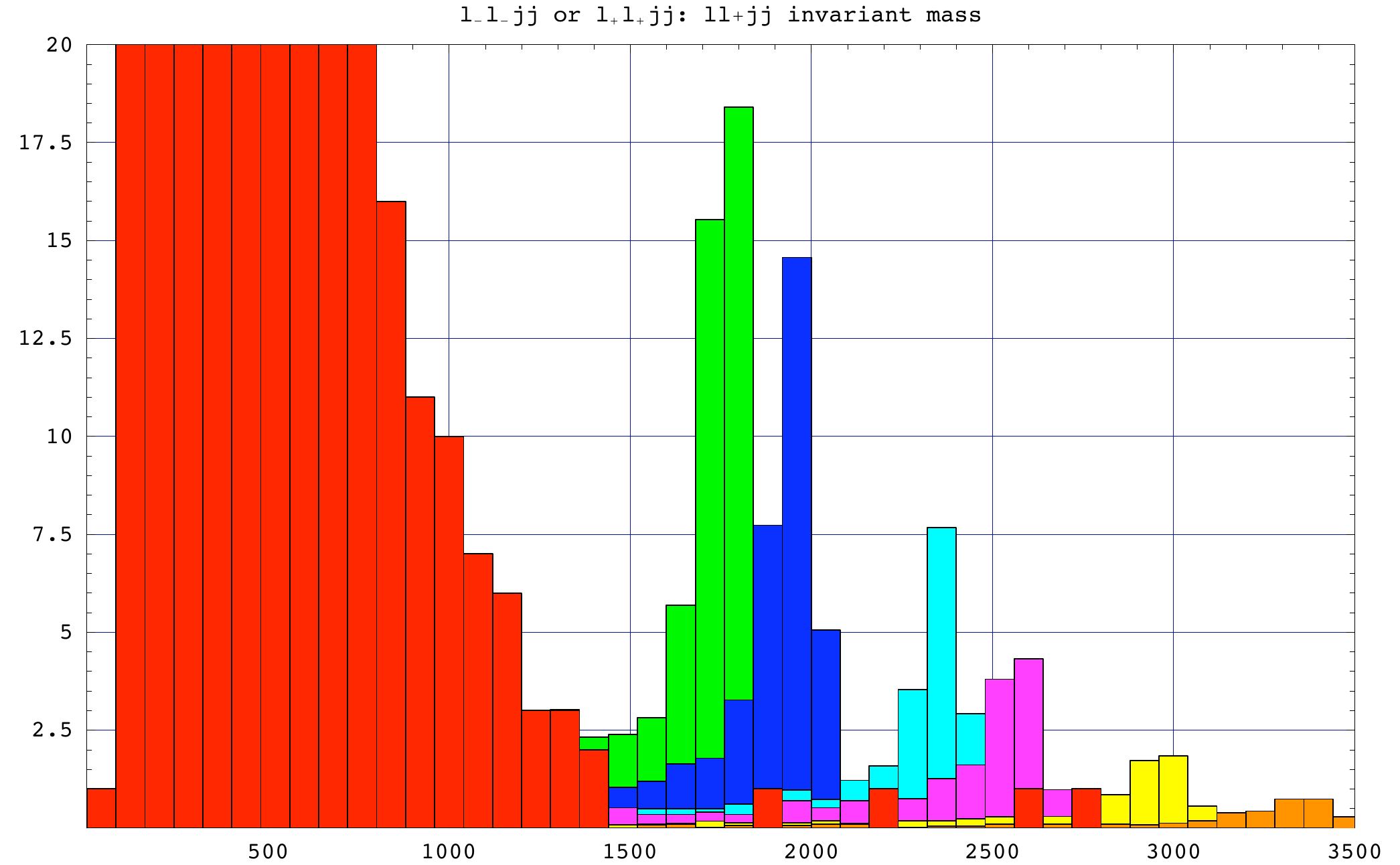} 
\caption{The number of events as a function of energy (GeV) for ${\rm L}=8{\rm fb} ^{-1}$ (courtesy of F. Nesti)
 where ${\rm M_R}$ (TeV) is  taken to be: $1.8;\,2,0;\,2.4;\,2.6;\,3,0;\,3.4$.}
 \label{dilep}
\end{center}
\end{figure*}

Once the right-handed gauge boson is produced, it will decay into a right-handed neutrino and a charged lepton.  The 
right-handed neutrino, being a Majorana particle, decays equally often into charged leptons or anti-leptons and jets.
This often confuses people for naively one argues that the production of a wrong sign lepton must be suppressed
by the mass of the right handed neutrino. True, but so does the production of the right sign lepton in its decay; this
is the usual time dilation. It is enough that $N$ is heavy enough as to decay into a lepton and two jets, and then the
above claim must be true.
In turn one has exciting events of same sign lepton pairs and two jets, as a clear signature of lepton number violation.
This is a collider analog of neutrino-less double beta decay, and it allows for the determination of $W_R$ mass as
shown in the Fig. \ref{dilep}.

  This offers 
\begin{list}{}{}
\item  a) direct test of parity restoration through a  discovery of $W_R$,
\item b) direct test of lepton number violation through a Majorana nature of $\nu_R$,
\item c) determination of $W_R$  and $N$ masses.
\end{list}
  A detailed study \cite{Ferrari:2000sp} concludes an easy  probe of $W_R$ up to 3.5 TeV and $\nu_R$ in 100 - 1000 GeV
     for integrated luminosity of 30 ${\rm fb} ^{-1}$. It needs a study of flavor dependence, i.e.  connection with LFV.
There have been recent claims of $M_R \gtrsim 4$ TeV \cite{Zhang:2007da} (or even $M_R \gtrsim 10$ TeV \cite{Xu:2009nt})
 in the minimal theory, but these limits depend on the definition of $L-R$ symmetry and its manifestness. Namely, this
 limit stems from CP violation which depends on the definition of $L-R$ symmetry.

Recall that $L-R$ symmetry can be P as in the original works or C as it happens in $SO(10)$. The authors of  \cite{Zhang:2007da, Xu:2009nt} use P, but it can be shown that in the case of C, the freedom in CP phases leaves only the CP conserving limit $M_R \gtrsim 2.4 \mbox{ TeV}$ \cite{Zhang:2007da}. This allows for both $W_R$ and the accompanying 
neutral gauge boson $Z_R$ to see seen at LHC.

   It is worth noting that the same signatures can be studied in the SM with $\nu_R$ \cite{Datta:1991mf}, but it 
   requires miraculous cancellations of large Dirac Yukawa couplings in order to keep neutrino masses small.
  When a protection symmetry is called for, one ends up effectively with lepton number conservation and the 
  phenomenon disappears \cite{Kersten:2007vk}. 
  
The $L-R$ theory possesses naturally also type II seesaw \cite{Mohapatra:1980yp}. The type II offers another  potentially interesting signature:
pair production of doubly charged Higgses  which decay into same sign lepton (anti lepton) pairs \cite{Han:2007bk}.  This can serve as
a determination of the neutrino mass matrix in the case when type I is not present or very small \cite{Kadastik:2007yd}.
It is worth commenting that the minimal supersymmetric left-right symmetric model \cite{Aulakh:1998nn} predicts 
doubly charged scalars at the collider energies \cite{Aulakh:1998nn} \cite{Chacko:1997cm} \cite{Aulakh:1997fq} even
for large scale of left-right symmetry breaking.
    
    This is all very nice, but the question is whether a low $L-R$ scale is expected or not. It is perfectly allowed, but 
not predicted. This theory can be embedded in $SO(10)$ grand unified theory, where $L-R$ symmetry becomes charge
conjugation and is a finite gauge transformation. The scale of $L-R$ breaking ends up being high though, either close
to $M_{GUT}$ in the supersymmetric version  \cite{Aulakh:2000sn}  \cite{Clark:ai}, or around $10^{10}$ GeV or so in the ordinary version \cite{Bajc:2005zf}. 

   We are faced then with a question: is there a simple predictive grand unified theory with  seesaw at LHC?
The answer is yes, a minimal extension of the original Georgi-Glashow theory \cite{Georgi:1974sy} , with an addition of an adjoint fermion
representation \cite{Bajc:2006ia}.

\section{Minimal non supersymmetric $SU(5)$}

  The minimal $SU(5)$ theory consists of:  $24_H+5_H$ Higgs multiplets, where $24_H$ is used to break 
  the original symmetry to the SM one, and $5_H$ completes the symmetry breaking; and the three generation of quarks
  and leptons $3(10_F+\overline{5}_F)$. The theory fails for two reasons:

\begin{itemize}
\item
gauge couplings do not unify

  $\alpha_2$ and $\alpha_3$  meet at about $10^{16}$ GeV 
(similar as in the MSSM), but $\alpha_1$ meets $\alpha_2$ too early, at $\approx 10^{13}$ GeV

\item
neutrinos remains massless as in the SM.

\end{itemize}
 The $d=5$ Weinberg operator for neutrino mass we started with is not enough: neutrino mass comes out too small
  ($\lesssim 10^{-4} eV$) since
 the cut-off scale M must be at least as large as $M_{GUT}$ due to $SU(5)$ symmetry. In any case, one must first make sure that the
 theory is consistent and the gauge couplings unify.

A simple extension cures both problems: add just one extra fermionic $24_F$  \cite{Bajc:2006ia}. This 
requires higher dimensional operators just as in the minimal theory, but can be made renormalizable as usual by
adding extra $45_H$ scalar \cite{Perez:2007rm}.

  Under $SU(3)_C\times SU(2)_W\times U(1)_Y$ the adjoint is decomposed as:
  $24_F=(1,1)_0+(1,3)_0+(8,1)_0+(3,2)_{-5/6}+(\bar 3,2)_{5/6}$.  The unification works as follows:
  triplet  fermion (like wino in MSSM) slows down $\alpha_2$ coupling without affecting $\alpha_1$. In order
  that they meet above $10^{15}$ GeV for the sake of proton's stability, 
  the triplet must be light, with a mass
  below TeV. Then in turn $\alpha_3$
  must be slowed down, which is achieved with an intermediate scale mass for the color octet in $24_F$ around
  $10^7$ GeV or so. 

  For a practitioner of supersymmetry, 
  the theory behaves  effectively as  the MSSM with a light wino,  gluino heavy ($10^7 \text{GeV}$), no Higgsino,
no sfermions (they are irrelevant for unification being complete representations). This
shows how splitting supersymmetry \cite{ArkaniHamed:2004fb} opens a Pandora's box of possibilities for unification.  
  The great success of low energy supersymmetry was precisely the prediction of gauge coupling unification
   \cite{Dimopoulos:1981yj} \cite{Ibanez:1981yh} \cite{Einhorn:1981sx} \cite{Marciano:1981un},
 ten years before the LEP confirmation of its prediction $\sin^2 \theta_W = .23$.  In 1981 when  it was thought that 
   $\sin^2 \theta_W = .21$,  this required asking for a heavy top quark, with $m_t \simeq 200$ GeV  \cite{Marciano:1981un}.
 Unlike the case of supersymmetry, where the scale was fixed by a desire for the naturalness  of the Higgs mass,  and then
 unification predicted, in this case the $SU(5)$ structure demands unification which in turn fixes the masses of the new 
 particles in $24_F$. The price is the fine-tuning of these masses, but a great virtue is the tightness of the theory: the low mass of the fermion triplet (and other masses) is a true phenomenological prediction not tied to a nice but imprecise notion of 
 naturalness.
  
  With the notation singlet $S=(1,1)_0$,  triplet $T=(1,3)_0$, it is evident that we have mixed Type I and Type III seesaw
\begin{eqnarray}
(M_\nu)^{ij}=v^2\left(
\frac{y_{T}^{i}y_{T}^{j}}{m_{T}}+
\frac{y_{S}^{i}y_{S}^{j}}{m_{S}}\right)\nonumber
\end{eqnarray}

An immediate consequence is 
one massless neutrino.  Thus one cannot have four generations in this theory, for
then all four neutrinos would be light which the Z decay width does not allow.  Since the triplet may
be out of LHC reach, seeing the fourth generation would serve an important test of a theory; it would
simply rule it out.

\subsection{ $T$ at LHC  }
\begin{figure*}
\begin{center}
\includegraphics[scale=0.48]{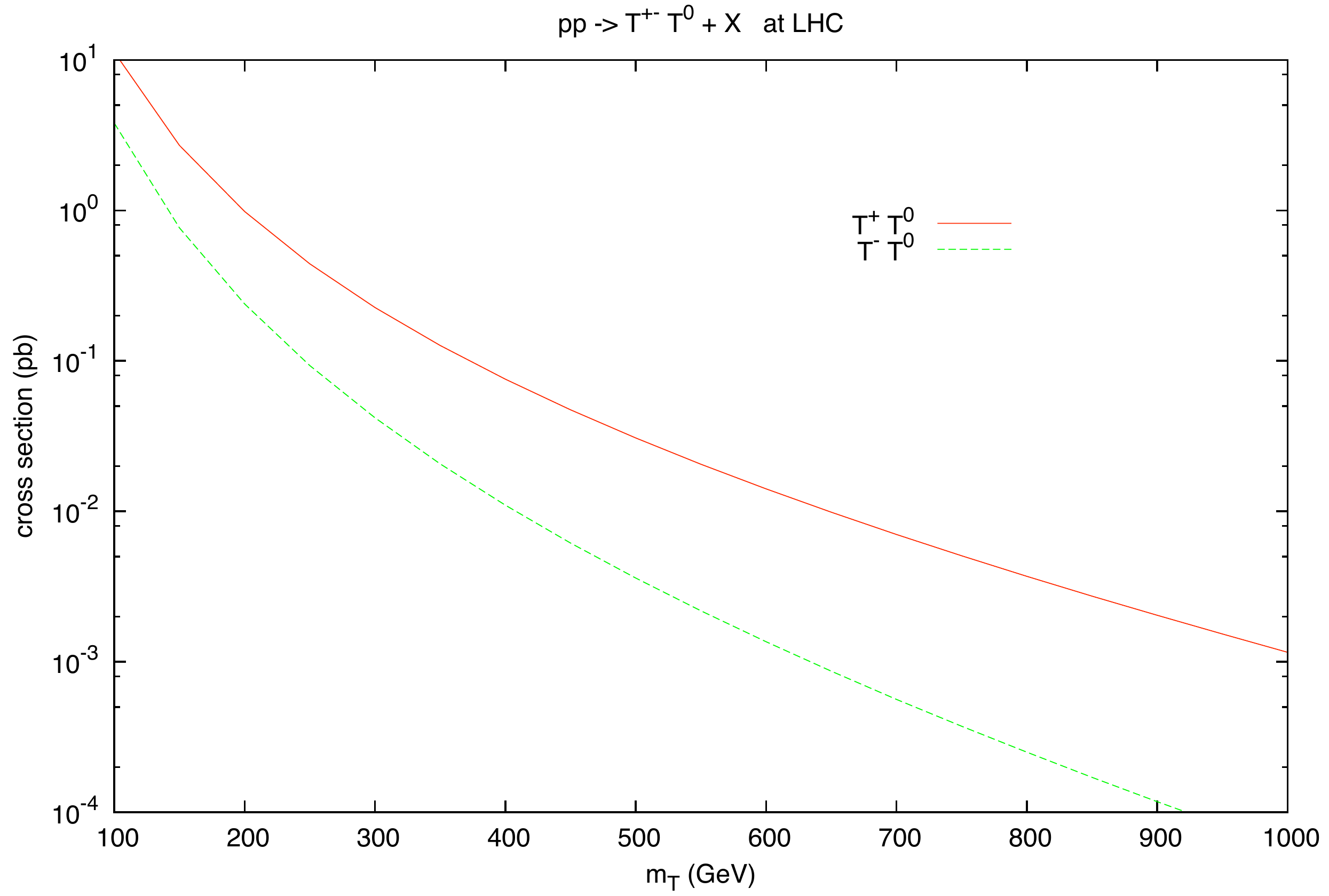} 
\caption{Cross section for the T pair production at LHC.} \label{pairprod}
\end{center}
\end{figure*}

 We saw that unification predicts the mass of the fermion triplet below TeV, and thus it becomes accessible to the colliders such as Tevatron and LHC.
 It can be produced through gauge interactions (Drell-Yan)
\begin{eqnarray}
pp\to W^\pm +X\to T^\pm T^0 +X\nonumber\\
pp\to (Z\,{\rm or}\, \gamma)+X\to T^+T^-+X\nonumber
\end{eqnarray}
with the cross section for the T pair production in Fig. \ref{pairprod}.

The best channel is like-sign dileptons + jets
\begin{eqnarray}
BR(T^\pm T^0\to l_i^\pm l_j^\pm +4{\rm jets})\approx\frac{1}{20}\times
\frac{|y_T^i|^2|y_T^j|^2}{(\sum_k|y_T^k|^2)^2}\nonumber
\end{eqnarray}

Same couplings $y_T^i$ contribute to $\nu$ mass matrix and $T$ decays, so that T decays can serve to probe the
neutrino mass matrix \cite{Bajc:2007zf} and the nature of the hierarchy of neutrino masses.

With proper cuts SM backgrounds appear under control  \cite{Franceschini:2008pz}.
With integrated luminosity of 10 ${\rm fb} ^{-1}$  one could find the fermionic triplet $T$ for $M_T$ up to about 400 GeV.

 The light triplet fermion also plays an important role in lepton flavor violation, especially in $\mu \to e$ conversion
 in nuclei,
 which is induced at the tree level and could be observed even for a triplet out of LHC reach \cite{Abada:2008ea}.

Before concluding, it should be mentioned that one can also add a 15-dimensional scalar as an alternative
of curing the minimal $SU(5)$ theory. This leads instead to the type II seesaw with possibly light  lepto-quarks and its own interesting phenomenology  \cite{Dorsner:2005ii}.

\section{Summary and Outlook}

I discussed here an  experimental probe of Majorana neutrino mass origin, both at colliders 
through the production of the same sign dileptons,
and a neutrino-less double beta decay.  A classical example is provided  by the $L-R$ symmetric theory that predicts
the existence of right-handed neutrinos and leads to the seesaw mechanism.
A TeV scale $L-R$ symmetry, as discussed here, would have spectacular signatures at LHC,
with a possible discovery of $W_R$ and $\nu_R$.  This offers a possibility of observing parity restoration and the Majorana nature of neutrinos. It is important to search for an underlying theory that predicts it naturally. For a recent attempt, see \cite{Dev:2009aw}. 

I have provided next an explicit example of a predictive grand unified theory: ordinary 
minimal $SU(5)$ with extra fermionic adjoint.   A weak fermionic triplet is predicted in the TeV range (type III seesaw) whose
decay is connected with neutrino mass.  This offers good chances for discovery at LHC with integrated luminosity of 10 ${\rm fb} ^{-1}$ for $M_T$ up to about 400 GeV.

One can also simply study the minimalist scheme of pure seesaw in the connection with the colliders. The type II and III
are naturally rather exciting from the experimental point of view, for the new states can be easily produced through the
gauge couplings. In the case of the type I it becomes a long shot,  since the Dirac Yukawas must be large and the smallness
of neutrino mass is then attributed to the cancellations. Strictly speaking that should not be called the seesaw whose name
was meant to indicate a natural smallness of neutrino mass after the heavy states are integrated out.  

In summary, I argued here that in spite the smallness of neutrino masses, the hope of probing their origin at
LHC is not just wishful thinking. Small Yukawa couplings are as natural as the large ones, and the low scale seesaw is perfectly
realistic, and even likely in the context of the $SU(5)$ grand unified theory.  There are other possible ways of 
having TeV scale seesaw,  as e.g. with mirror leptons \cite{Hung:2006ap} and in the case of dynamical symmetry
breaking \cite{Appelquist:2002me}.
  
 \section{Acknowledgements}

  I wish to thank Steven Parke and other organizers of Neutrino 08 for a great conference in a beautiful setting; and
  Daniel Denegri, Ivica Puljak and other organizers of Physics at LHC - 2008 for yet another excellent conference 
  in the most beautiful town in the world (\v{c}a je pusta Londra kontra Splitu gradu). 
  I am
  deeply grateful to my collaborators on the topics covered above  Abdesslam  Arhrib,  Charan Aulakh,  Dilip Ghosh, Tao Han, 
  GuiYu Huang,  Alejandra Melfo, Miha Nemev\v{s}ek, Fabrizio Nesti, Ivica Puljak, Andrija Ra\v{s}in (Andrija, vrati se, sve 
  ti je oprosteno), Francesco Vissani and especially Borut Bajc. 
  I acknowledge with great pleasure  my collaboration with Wai-Yee Keung
   on the issue of lepton number violation at colliders. I  am grateful to Enkhbat Tsedenbaljir for numerous useful discussions. 
  Thanks are also due to Alejandra Melfo and Vladimir Tello for their help with this manuscript, and Borut Bajc,  Miha 
  Nemev\v{s}ek and Enkhbat Tsedenbaljir for careful reading of the manuscript.
   This work was partially supported by the EU FP6 Marie Curie Research and Training Network "UniverseNet" (MRTN-CT-2006-035863).

\end{document}